\documentclass[twocolumn,notitlepage]{revtex4}
\usepackage{amsmath,amssymb}
\usepackage{bm}
\usepackage{epsfig}
\usepackage{graphicx}
\usepackage{color}
\usepackage[english]{babel}

\newcommand{\erfc}{\mathop{\rm erfc}\nolimits}

\renewcommand{\Re}{\mathop{\rm Re}}

\begin{document}
%
%
%\newcount\timehh  \newcount\timemm
%\timehh=\time \divide\timehh by 60
%\timemm=\time
%\count255=\timehh\multiply\count255 by -60 \advance\timemm by \count255

\title{Spin noise of localized electrons:\\ Interplay of hopping and hyperfine interaction}
\author{M.M. Glazov}
\affiliation{Ioffe Institute, Russian Academy of Sciences, 194021,
St.-Petersburg, Russia}
%
%\date{\today, file = \jobname.tex, printing time = \number\timehh\,:\,\ifnum\timemm<10 0\fi \number\timemm}

\begin{abstract}
The theory of spin fluctuations is developed for an ensemble of localized electrons taking into account both hyperfine interaction of electron and nuclear spins and electron hopping between the sites. The analytical expression for the spin noise spectrum is derived for arbitrary relation between the electron spin precession frequency in the field of the  nuclear fluctuations and the hopping rate. An increase in the hopping rate results in the drastic change in the spin noise spectrum. The effect of an external magnetic field is briefly addressed.
\end{abstract}
\maketitle

%\section{Introduction}\label{sec:intro}

\emph{Introduction.} The rise of the spin noise spectroscopy technique has opened new prospects for the spin dynamics studies, particularly, in close-to-equilibrium conditions~\cite{aleksandrov81,Crooker_Noise,Zapasskii:13,Oestreich:rev}. In this technique the spectrum of spin fluctuations in a media and, correspondingly, its spin susceptibility  is measured by fluctuations of spin-Kerr or Faraday effects. The method first suggested and demonstrated for atomic systems has found wide applications in semiconductor spintronics. It is  of special importance for the systems with localized charge carriers, such as donor-bound electrons in bulk semiconductors, quantum wells with low-density electron gas, and quantum dots~\cite{crooker2012,PhysRevLett.112.156601,PhysRevB.89.081304}.  In these systems the spin-orbit coupling effects are quenched and the spin relaxation and decoherence are mainly caused by the hyperfine interaction with nuclear spins~\cite{Dzhioev02,merkulov02,PhysRevLett.88.186802,0268-1242-23-11-114009}. The theory of  spin noise in ensembles of localized electrons or holes has been developed in Ref.~\cite{gi2012noise}, see also Ref.~\cite{crooker2012},  the quadrupolar effects have been addressed in Ref.~\cite{PhysRevLett.109.166605}. Reference~\cite{PhysRevB.89.045317} studied the effects of nuclear spin dynamics in the spin noise, and  the effect of exchange interaction on spin fluctuations has beed considered in Ref.~\cite{gis2014noise}. It has been assumed that the charge carriers remain at the same site during all the time relevant for the spin dynamics. In other words, the electron correlation time i.e. its lifetime at a given donor or quantum dot, $\tau_c$, was assumed to be infinite.

This assumption of infinite correlation time is, however, not universal. The experiments carried out on bulk semiconductors demonstrate finite values of $\tau_c$, see, e.g., Refs.~\cite{Dzhioev:2002kx,Dzhioev02} and Ref.~\cite{0268-1242-23-11-114009} for review. Depending on the donor density and the temperature, presence of free carriers or optical excitation, the correlation time $\tau_c$ can be either shorter or longer than the electron spin precession period in the field of nuclear spin fluctuation. In these two limits the electron spin noise spectra are expected to be qualitatively different. If $\delta_e\tau_c \gg 1$, where $\delta_e$ is the characteristic electron spin precession frequency in the field of nuclear fluctuation, the spectrum is described by the theory of Ref.~\cite{gi2012noise} and contains at frequencies $\omega\geqslant 0$ two peaks. One of the peaks is related with spin precession in the static fields of nuclear fluctuations. The other one is centered at $\omega=0$ and caused by the electron spin components parallel to the nuclear fields, which are conserved during the spin precession. If, by contrast, $\delta_e \tau_c \ll 1$, the noise spectrum should have a single peak centered at zero frequency due to the dynamical averaging of the nuclear fields in electrons hopping~\cite{dp74,kkm_nucl_book}. In this regime the spin relaxation substantially slows down. Hence, the effect of hopping on electron spin dynamics has attracted recently a significant interest  related to the search of  systems with ultralong spin relaxation times based both on conventional~\cite{kikkawa,Dzhioev02} and organic semiconductors~\cite{dediu}. The spin dynamics in the hopping regime with allowance for the spin-orbit interaction and neglecting nuclear effects was addressed e.g. in Refs.~\cite{shklovskii:193201,lyub:07:eng,PhysRevLett.108.016601}. An interplay of the hyperfine interaction and carrier hopping in spin relaxation and diffusion was studied in Refs.~\cite{PhysRevLett.102.156604,PhysRevLett.110.176602,roundy2014} for molecular systems, and in old works on muon spin dynamics, see Ref.~\cite{muon} for review. Despite these in-depth studies, a simple model for the arbitrary $\delta_e\tau_c$ suitable for direct comparison with the spin noise experiments is absent to the best of our knowledge.

The aim of the present work is to trace the evolution of the electron spin noise spectrum as a function of the correlation time. We present simplest model of electron hopping between different sites each of those is characterized by a static nuclear fluctuation. Under the assumption of equal hopping rates we derive analytical expression for the spin noise spectrum valid for the arbitrary value of the $\delta_e\tau_c$. An external magnetic field is also included into consideration and its effects are discussed.

%\section{Model and formalism}\label{sec:model}

\begin{figure}[t]
\includegraphics[width=0.85\linewidth]{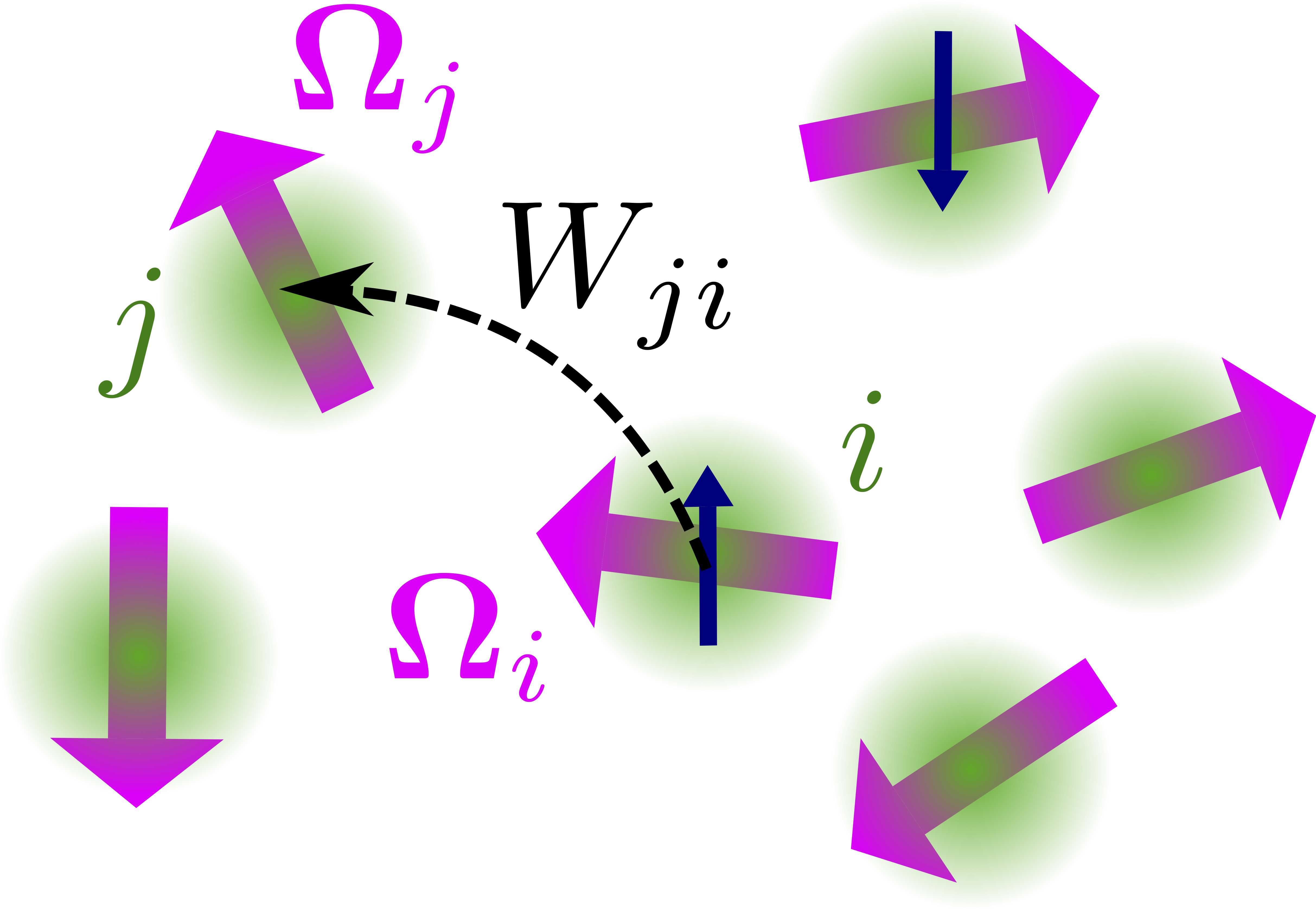}
\caption{Sketch of the ensemble of localized states. Thick magenta arrows denote nuclear spin fluctuations $\bm \Omega_i$, thin dark blue arrow denotes electron spin, dashed arrow shows the electron jump from the occupied  site $i$ to the empty site $j$.} \label{fig:scheme}
\end{figure}

\emph{Model and formalism.} We consider $N$ localization sites for an electron with nuclear fluctuations, which result in  static effective fields acting on electron spins with the precession frequencies $\bm \Omega_i$, $i=1, 2,\ldots, N$. Such a model can be relevant e.g. for partially compensated bulk semiconductors, quantum wells with low-density electrons or ensembles of quantum dots with average occupancy per dot being smaller than $1$. The single electron spin density matrix can be parametrized by $\bm S_i$, the spin pseudovectors at the given site, which obey the following set of Bloch equations
\begin{multline}
\label{set}
\frac{\mathrm d \bm S_i}{\mathrm dt} = \bm \Omega_i \times \bm S_i + \sum_j [W_{ij} \bm S_j - W_{ji}\bm S_i] - \nu_s \bm S_i, \\ i=1,\ldots, N,~ j=1,\ldots, N.
\end{multline}
Here $W_{ij}$ are the hopping rates from site $j$ to site $i$ and $\nu_s$ is the spin relaxation rate unrelated with hyperfine interaction and hopping. The system is illustrated in Fig.~\ref{fig:scheme}. Spin rotation at the jumps due to the spin-orbit coupling is disregarded; a more general situation is addressed in Ref.~\cite{PhysRevLett.110.176602} for organic semiconductors.  We neglect exchange interaction between the electrons as well~\cite{gis2014noise}.
Set of Eqs.~\eqref{set} is applicable assuming that characteristic hopping times $\sim W_{ij}^{-1}$ are much smaller as compared with the time-scale of nuclear spin dynamics, otherwise  temporal variations of $\bm \Omega_i$ should be taken into account~\cite{merkulov02,PhysRevLett.97.037204,PhysRevB.84.155315,PhysRevB.89.045317}.

We introduce the autocorrelation function of the electron spin components, $\langle S_{\alpha}(t'+t)S_{\beta}(t')\rangle$, where $S_\alpha = \sum_i S_{i,\alpha}$, $\alpha,\beta = x,y,z$ enumerate Cartesian components, the angular brackets denote averaging over $t'$ at fixed  $t$. Our aim is to calculate the spin noise power spectrum 
\begin{equation}
\label{sns}
(S_\alpha S_\beta)_\omega = \int_{-\infty}^\infty \langle S_\alpha(t'+t)S_\beta(t')\rangle \exp{(\mathrm i \omega t)}\mathrm dt.
\end{equation} To that end we introduce a set of correlation functions 
\begin{equation}
\label{cor:def}
\mathcal C^{(ij)}_{\alpha\beta} \equiv  \int_{0}^\infty  \langle S_{i,\alpha}(t'+t)S_{j,\beta}(t')\rangle \exp{(\mathrm i \omega t)}\mathrm dt',
\end{equation}
 which according to the general principles of fluctuation theory~\cite{ll10_eng,gi2012noise,gis2014noise} satisfy the following equations
 \begin{multline}
\label{set:C}
\left(-\mathrm i \omega + \nu_s + \sum_{j'} W_{j'i}\right) \mathcal C^{(ij)}_{\alpha\beta} +\sum_{\alpha'\beta'}\epsilon_{\alpha\alpha'\beta'} \Omega_{i\beta'}  \mathcal C^{(ij)}_{\alpha'\beta}\\
 - \sum_{j'} W_{ij'} \mathcal C^{(j'j)}_{\alpha\beta} = \frac{1}{4N}\delta_{ij}\delta_{\alpha\beta}.
\end{multline}
Here $\epsilon_{\alpha\beta\gamma}$ is the third-rank antisymmetric tensor, $\delta_{\alpha\beta}$ is the Kronecker symbol. In derivation of Eq.~\eqref{set:C} we assumed that the equilibrium occupancies of all sites are small and equal to each other. In the absence of external magnetic field the spin noise spectrum is isotropic $(S_x^2)_\omega = (S_y^2)_\omega = (S_z^2)_\omega$. For definiteness we assume that the probe beam used to measure the spin noise propagates along $z$-axis and, correspondingly, calculate $(S_z^2)_\omega =2\Re\{ \sum_{ij} \mathcal C^{ij}_{zz}\}$. Solution of Eqs.~\eqref{set:C} allows one to obtain the spin noise spectrum.

 In order to derive a closed form analytical expression for the spin noise power spectrum we assume that all hopping rates are the same, $W_{ij} = W_0/N$, where $W_0$ is related with electron correlation time, $\tau_c$, on a site as $W_0 = 1/\tau_c$; a brief discussion of the applicability of this model is given below. Summing up Eqs.~\eqref{set:C} over site $j$ and introducing the notations 
 $$\bm{\mathcal S}^i_\beta = \left(\sum_j \mathcal C^{(ij)}_{x\beta}, \sum_j \mathcal C^{(ij)}_{y\beta},\sum_j \mathcal C^{(ij)}_{z\beta}\right),$$ 
 we arrive at
  \begin{equation}
\label{set:C1}
\left(-\mathrm i \omega + \nu_s + W_0\right) \bm{\mathcal S}^i_\beta  +\bm{\mathcal S}^i_\beta \times \bm  \Omega_{i}=\frac{1}{4}\bm g_\beta  + \frac{W_0}{N} \sum_i  \bm{\mathcal S}^i_\beta,
\end{equation}
with $\bm g_\beta =(\delta_{x\beta},\delta_{y\beta},\delta_{z\beta})$. Equation~\eqref{set:C1} can readily be solved and the spin noise power spectrum (per electron) is given by~\footnote{{The Fourier transform of Eq.~\eqref{snoise:analyt} provides the spin dynamics after pulsed excitation.}}
\begin{equation}
\label{snoise:analyt}
(S_z^2)_\omega = \frac{\tau_\omega}{4} \frac{\mathcal A(\tau_\omega)}{1-W_0\tau_\omega \mathcal A(\tau_\omega)} + {\rm c.c.},
\end{equation}
where c.c. denotes the complex conjugate,
\begin{equation}
\label{tau:omega}
\tau_{\omega}^{-1} = \nu_s + W_0 - \mathrm i \omega,
\end{equation}
and 
\begin{equation}
\label{A}
\mathcal A(\tau_\omega) = \frac{1}{N}\sum_i\frac{1+\Omega_{iz}^2\tau_\omega^2}{1+\Omega_i^2\tau_\omega^2}.
\end{equation}
In derivation of Eqs.~\eqref{snoise:analyt} and \eqref{A} we assumed that the nuclear fields are isotropically distributed. In the case of a macroscopic system where $N\to \infty$ and the nuclear fields $\bm \Omega_i$ being sums of random contributions of individual nuclei obey the Gaussian distribution in the form $\mathcal F(\bm \Omega) = (\sqrt{\pi} \delta_e)^{-3} \exp{(-\Omega^2/\delta_e^2)}$~\cite{gi2012noise} we obtain
\begin{equation}
\label{Aa}
\mathcal A(\tau_\omega) =  \frac{1}{3} + \frac{4}{3(\delta_e\tau_\omega)^2} - \frac{4\sqrt{\pi}\, \mathrm e^{1/\delta_e^2\tau^2_\omega}}{3(\delta_e\tau_\omega)^3}  \erfc{(1/\delta_e\tau_\omega)}.
\end{equation}
Here $\erfc{(y)} = 1- (2/\sqrt{\pi})\int_0^y \exp{(-x^2)} \mathrm dx$ is the complementary error function. Equations~\eqref{snoise:analyt} and \eqref{Aa} describe the spin noise spectrum for arbitrary relation between $W_0$ and $\delta_e$, i.e. for arbitrary value of $\delta_e\tau_c$. Note that particular form of $\mathcal A(\tau_\omega)$ is determined by the shape of the nuclear fields distribution $\mathcal F(\bm \Omega)$.
 For GaAs-based systems characteristic values of $\delta_e^{-1} \sim 1\ldots 10$~ns and electron correlation time at a donor $\tau_c$ ranges from $\sim 10$~ps to $\sim 10$~ns depending on the donor density, temperature, etc., see Refs.~\cite{0268-1242-23-11-114009,kkm_nucl_book} for review. Weaker hyperfine coupling with $\delta_e^{-1} \gtrsim 40$~ns can be realized in ZnSe with Fluorine donors where donor-bound electron interacts with smaller number of nuclear spins~\cite{greilich:Fl} 
 
%\section{Results and discussion}\label{sec:res}

\emph{Results and discussion.} To analyse the spin noise spectrum for the arbitrary $W_0/\delta_e$ it is instructive to address the limits of long and short correlation times, respectively. Hereinafter we assume that the spin relaxation rate unrelated with the hyperfine interaction and electron jumps, $\nu_s$, is much smaller than $\delta_e$. In the limit of long correlation time where $W_0 \ll \delta_e$ and for $\omega \gg W_0$, $\nu_s$, one can neglect $W_0$ and $\nu_s$ in $\tau_\omega$ and made use of the fact that for real $y$ $\Re\{\erfc{(\mathrm i y)}\} =1$, yielding in agreement with Ref.~\cite{gi2012noise}
\begin{subequations}
\label{long}
\begin{equation}
\label{prec:long}
(S_z^2)_\omega = \frac{2\sqrt{\pi}}{3\delta_e^3}\omega^2\exp{(-\omega^2/\delta_e^2)}, \quad \omega,\delta_e \gg W_0, \nu_s.
\end{equation}
It is the precession peak in the spin noise spectrum whose shape is determined by the distribution function of the nuclear field absolute values $4\pi \Omega^2 \mathcal F(\Omega)$~\cite{gi2012noise}. At low frequencies $\omega \ll \delta_e$ it is enough to put $\mathcal A(\tau_\omega)=1/3$ with the result
\begin{equation}
\label{T1:long}
(S_z^2)_\omega = \frac{1}{6} \frac{\Gamma_0}{\Gamma_0^2+\omega^2}, \quad \omega, W_0, \nu_s \ll \delta_e,
\end{equation}
\end{subequations}
where $\Gamma_0 = \nu_s + 2W_0/3$. Equation \eqref{T1:long} describes zero-frequency (or $T_1$) peak in the spin noise spectrum caused by the spin components conserved during the precession~\cite{gi2012noise}. Effective longitudinal relaxation time $T_1 = 1/\Gamma_0$ is limited by the fastest of either the spin-flip process ($\sim \nu_s^{-1}$) or the hopping ($\sim W_0^{-1}$). In the latter case when electron leaves given site $i$, the components of spin {parallel to} $\bm \Omega_i$ start to precess in the field of nuclear field fluctuation $\bm \Omega_j$ on the new site (unless, of course, $\bm \Omega_i \parallel \bm \Omega_j$) and this electron stops to contribute to zero-frequency peak.

In the opposite limit of short correlation time where $W_0 \gg \delta_e$ one needs to develop $\mathcal A(\tau_\omega)$ in power series in $\delta_e$ with the result $\mathcal A(\tau_\omega) =1- (\delta_e\tau_\omega)^2$ and 
\begin{equation}
\label{short}
(S_z^2)_\omega = \frac{1}{2} \frac{\Gamma_1}{\Gamma_1^2+\omega^2}, \quad \delta_e \ll \omega \ll W_0,
\end{equation}
where $\Gamma_1 = \delta_e^2/W_0 +\nu_s$. In this limit the spin noise spectrum consists of the zero-frequency peak with the width decreasing with an increase in the hopping rate. This is typical for the motional narrowing regime realized for short correlation time, $\delta_e\tau_c \ll 1$, where the electron spin rotation angle during $\tau_c$ is small~\cite{dp74,kkm_nucl_book}. Note, that a  very long-time spin dynamics and, correspondingly, very low-frequency asymptotic of the spin noise can deviate from simple Lorentzians described by Eqs.~\eqref{T1:long}, \eqref{short} due to (i) nuclear spin dynamics and (ii) non-trivial hopping effects, e.g., particle returns, disregarded here~\cite{merkulov02,PhysRevLett.97.037204,PhysRevB.84.155315,PhysRevB.89.045317,PhysRevLett.110.176602,roundy2014}. 
 
\begin{figure}[t]
\includegraphics[width=\linewidth]{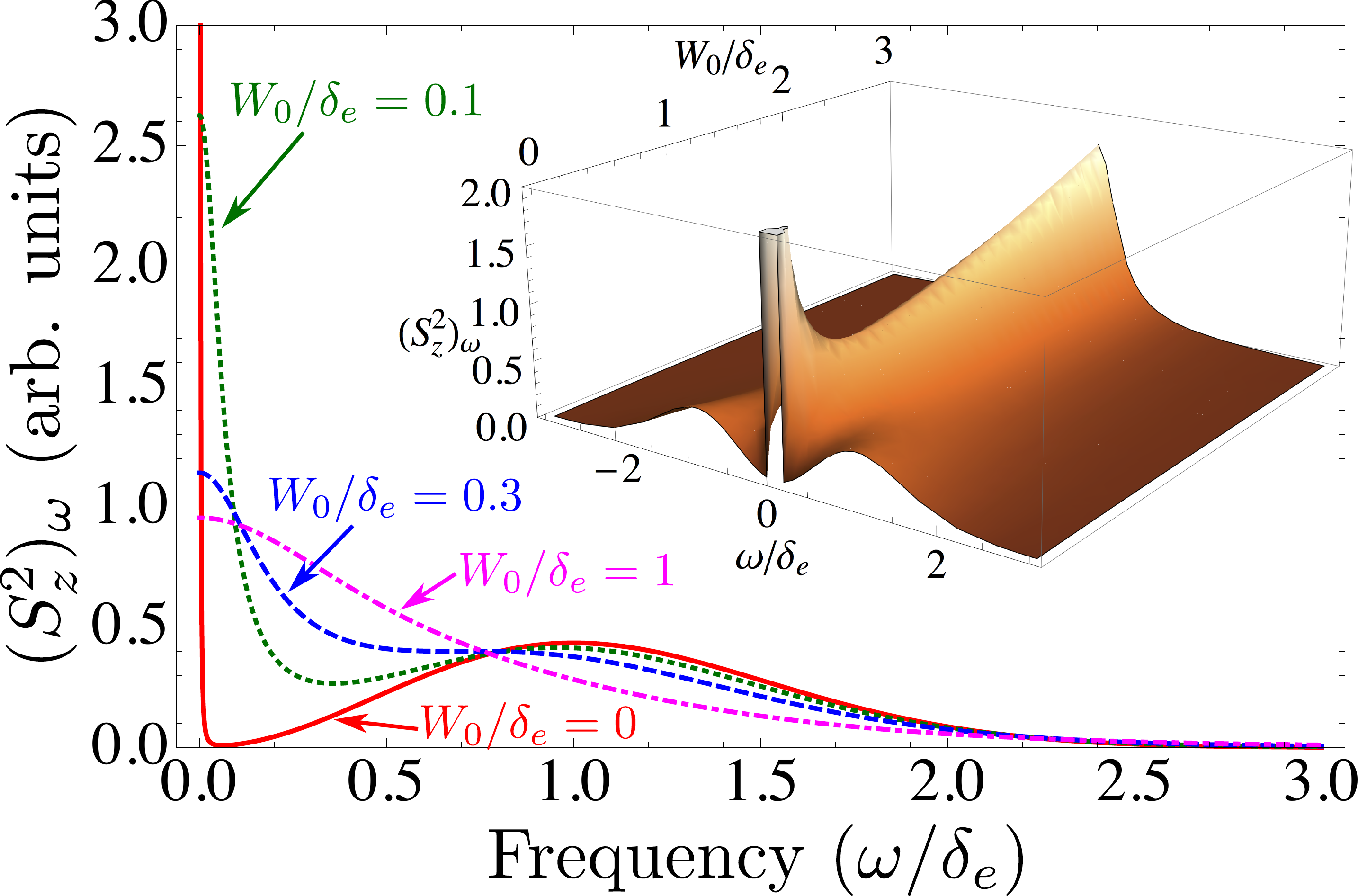}
\caption{Electron spin  noise power spectrum as a function of frequency $\omega/\delta_e$ for different hopping rates (marked at each curve) calculated after Eqs.~\eqref{snoise:analyt}, \eqref{tau:omega}, \eqref{Aa} at $\nu_s /\delta_e = 10^{-4}$. Inset shows three-dimensional plot. } \label{fig:corr}
\end{figure} 
 
Figure~\ref{fig:corr} shows the electron spin noise power spectrum as function of the hopping rate $W_0$ and demonstrates the transition between the limits of long and short correlation times. For the long correlation time where the jumps between the sites are suppressed the spectrum demonstrates two-peak structure~\cite{gi2012noise}. An increase in the hopping rate results in the broadening of the zero-frequency peak and suppression of spin precession peaks. If $W_0/\delta_e \sim 1$ then zero-frequency and precession peaks merge and further increase in the hopping rate $W_0$ results in an increase in the height and the decrease in the width of the spin noise peak due to the dynamical averaging of the nuclear fields.

Now we briefly consider the electron spin fluctuations in the presence of an external magnetic field characterized by the electron spin precession frequency $\bm \Omega_B$. We consider two situations of (i) longitudinal field (parallel to the probe beam propagation axis), $\bm \Omega_B\parallel z$, and (ii) transverse field, $\bm \Omega_B \perp z$. It is noteworthy that Eqs.~\eqref{set}, \eqref{set:C}, and \eqref{set:C1} hold with the replacement 
\begin{equation}
\label{replacement}
\bm \Omega_i \to \bm \Omega_i + \bm \Omega_B.
\end{equation}
in the presence of arbitrary magnetic field provided that the field neither affects hopping rates $W_{ij}$ and nor leads to a significant equilibrium spin polarization.  Moreover, in the particular case (i) where $\bm \Omega_B\parallel z$ the spin noise spectrum is, as before, given by  Eqs.~\eqref{snoise:analyt}, \eqref{tau:omega}, \eqref{A} provided Eq.~\eqref{replacement} is taken into account in evaluation of sum in Eq.~\eqref{A}. The longitudinal magnetic field thus reduces the effect of nuclear fluctuations on $z$ spin component fluctuations suppressing (in the limit of long correlation time $W_0 \ll \delta_e$) the precession peak, shifting it to the higher frequencies $\omega \sim \Omega_B$, and enhancing zero-frequency peak. Indeed, provided that $\Omega_B \gg \delta_e, W_0$ the function $\mathcal A(\tau_\omega) \approx 1 - \delta_e^2/\Omega_B^2$ and the spin noise spectrum has a single-peak form of Eq.~\eqref{short} with $\Gamma_1 =  W_0 \delta_e^2/\Omega_L^2 + \nu_s$. Due to rapid Larmor precession of the transverse spin components the nuclear fields dynamically average-out and their effect on electron spin noise becomes reduced~\cite{Dzhioev:2002kx}.

\begin{figure}[t]
\includegraphics[width=\linewidth]{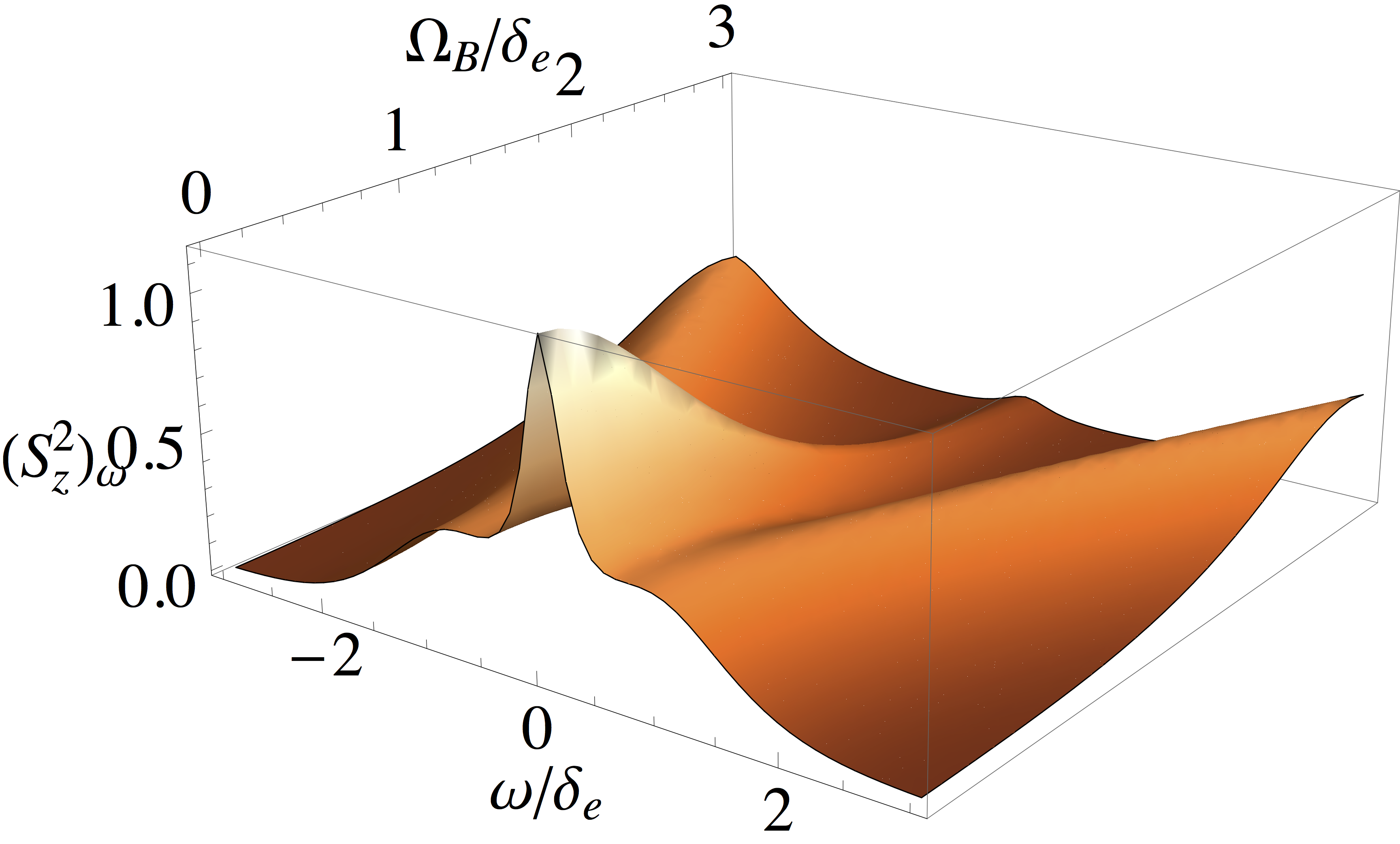}
\caption{Electron spin  noise power spectrum (arb. units) as a function of frequency $\omega/\delta_e$ and magnetic field $\Omega_B/\delta_e$ calculated after Eqs.~\eqref{tau:omega}, \eqref{snoise:analyt:B}, and \eqref{AB} at $W_0/\delta_e=0.3$ and $\nu_s /\delta_e = 10^{-4}$.} \label{fig:B}
\end{figure}  
 
The description of the transverse field effect is more complicated. Assuming that $\bm \Omega_{B}\parallel x$-axis,  we obtain instead of   Eq.~\eqref{snoise:analyt}
\begin{equation}
\label{snoise:analyt:B}
(S_z^2)_\omega = \frac{\tau_\omega}{4} \frac{ \mathcal A (\tau_\omega)[1-W_0\tau_\omega \mathcal A (\tau_\omega)] - (W_0\tau_\omega)^2 \mathcal B^2(\tau_\omega)}{[1-W_0\tau_\omega \mathcal A (\tau_\omega)]^2+(W_0\tau_\omega)^2 \mathcal B^2(\tau_\omega)} + {\rm c.c.},
\end{equation}
where $\mathcal A(\tau_\omega)$ is given by Eq.~\eqref{A} and $\mathcal B(\tau_\omega)$ is defined by
\begin{equation}
\label{AB}
\mathcal B(\tau_\omega) = \frac{1}{N}\sum_i\frac{\Omega_{xi}\tau_\omega}{1+\Omega_i^2\tau_\omega^2}.
\end{equation}
An increase in the  field $\Omega_B$ up to $\sim \delta_e$ results (in the limit of long correlation time $\delta_e\tau_c \gg 1$) in the suppression of zero-frequency peak and enhancement of the precession peak. Further increase in $\Omega_B$ results in the shift of the precession peak to the frequency $\omega \sim \Omega_B$. In the limit of short correlation time the spin precession peak shifts following the Larmor frequency $\Omega_B$. The transformation of the spin noise spectrum with an increase in  the transverse magnetic field is illustrated in Fig.~\ref{fig:B} for intermediate value of hopping rate $W_0/\delta_e=0.3$. The figure clearly shows the suppression of the zero-frequency peak and evolution of the spin precession peaks from hyperfine-induced satellites of zero-frequency peak to the peaks at the Larmor frequency.

It is noteworthy that in real disordered systems, e.g. bulk semiconductors with donor-bound electrons where the hopping is realized via electron-phonon interaction, the rates $W_{ij}$ are distributed over a wide range of values due to a spread of the hopping distances~\cite{ES:book}. By analogy with Refs.~\cite{Zvyagin,lyub:07:eng,gis2014noise} one can describe such a system as a set of clusters, i.e. groups of donors containing ionized ones where the hopping rates are large compared with spin precession rate in the nuclear field, $W_{ij}>\delta_e$~\footnote{Alternatively, the distribution of the hopping rates can be introduced by methods of Refs.~\cite{PhysRevLett.110.176602,roundy2014,muon}.}. Hence, in clusters the dynamical averaging of the nuclear fields is realized and the spin  noise spectrum in the absence of magnetic field has a single peak, Eq.~\eqref{short}, while remaining single sites provide two-peak contribution to the spin noise spectrum. The resulting spectrum has form similar to the those shown in Fig.~\ref{fig:corr} and transforms from a two-peak to a single peak shape with an increase in the hopping rates $W_{ij}$. In quantum dot ensembles an inhomogeneous broadening makes important effects on the observed spin noise spectra~\cite{gi2012noise,inhom}. Particularly, if typical electron energy change at the hopping exceeds the sensitivity band of spin Kerr/Faraday effect, then after the hopping it stops contributing to the measured signal. In this limit and at $\delta_e\tau_c \gg 1$, the hopping simply contributes to the spin decay rate $\nu_s \to \nu_s + W_0$. In the opposite limit of small energy transfer at the hopping the above theory is fully applicable.

%\section{Conclusions}

\emph{Conclusions.} To conclude, we have proposed the analytical model of spin fluctuations in an ensemble of localized electrons. The model takes into account an interplay of the  electron hopping between the sites and the electron spin precession in the random field of lattice nuclei. The transition between the weak hopping or long correlation time regime and the fast hopping and short correlation time regime has been analyzed. In the former limit the spin noise spectrum has a two-peak shape revealing the distribution of nuclear fields, while in the latter limit the noise spectrum has a single peak due to dynamical averaging of the fields. The effects of longitudinal and transverse magnetic fields have been also investigated. The developed theory can be also relevant for the nanowire systems where effects of spin-orbit coupling are partially suppressed leading to the features in the low-frequency spin noise spectra~\cite{PhysRevLett.107.156602}, while localization and hyperfine interaction may play important role~\cite{PSSR:PSSR201206257}. The obtained expressions can be applied for the ``express analysis'' of the spin noise spectra of localized electrons both in conventional and in organic semiconductors making it possible to extract characteristic nuclear field, $\delta_e$, and its correlation time $\tau_c = 1/W_0$.

%\acknowledgements

\emph{Acknowledgements.} Author is grateful to E.Ya. Sherman and D.S. Smirnov for valuable discussions. This work was partially supported by Russian Science Foundation (\# 14-12-01067), RF President Grant MD-5726.2015.2 and Dynasty Foundation.

\end{document}